\documentclass[acmlarge]{acmart}
\makeatletter
\newcommand{\myconfshort}{\acmConference@shortname}
\newcommand{\myconffull}{\acmConference@name}
\newcommand{\myconfdate}{\acmConference@date}
\newcommand{\myconfloc}{\acmConference@venue}
\AtBeginDocument{
  \fancypagestyle{firstpagestyle}{
    \fancyhead{}%
    \fancyfoot[C]{}%
  }
  \fancyhf{}
  \fancyhead[LO]{\@headfootfont\shorttitle}%
  \fancyhead[RE]{\@headfootfont\@shortauthors}%
  \fancyhead[LE]{\@headfootfont\footnotesize \myconfshort, \myconfdate, \myconfloc}%
  \fancyhead[RO]{\@headfootfont\footnotesize \myconfshort, \myconfdate, \myconfloc}%
  \fancyfoot[C]{}%
}
\makeatother
\AtBeginDocument{%
  }
\usepackage{longtable}
\usepackage{soul}

\copyrightyear{2026}
\acmYear{2026}
\setcopyright{cc}
\setcctype{by}
\acmConference[FAccT '26]{The 2026 ACM Conference on Fairness, Accountability, and Transparency}{June 25--28, 2026}{Montreal, QC, Canada}
\acmBooktitle{The 2026 ACM Conference on Fairness, Accountability, and Transparency (FAccT '26), June 25--28, 2026, Montreal, QC, Canada}
\acmDOI{10.1145/3805689.3812421}

\acmISBN{979-8-4007-2596-8/2026/06}
\acmConference[FAccT '26]{The 2026 ACM Conference on Fairness, Accountability, and Transparency}{June 25--28, 2026}{Montreal, Canada}

\begin{document}

\title{Open AI in the Wild: Adoption and Adaptation of Open Models on r/LocalLLaMA}

\author{Woohyeuk Lee}
\email{woohyeuk.lee@utexas.edu}
\orcid{0009-0003-8688-976X}
\affiliation{%
  \institution{University of Texas at Austin, School of Information}
  \city{Austin}
  \state{Texas}
  \country{USA}
}

\author{James Howison}
\orcid{0000-0002-5702-149X}
\affiliation{%
  \institution{University of Texas at Austin, School of Information}
  \city{Austin}
  \state{Texas}
  \country{USA}
}
\email{jhowison@ischool.utexas.edu}

\author{Min Kyung Lee}
\orcid{0000-0002-2696-6546}
\authornote{Both senior authors contributed equally.}
\affiliation{%
  \institution{University of Texas at Austin, School of Information}
  \city{Austin}
  \state{Texas}
  \country{USA}
}
\email{minkyung.lee@austin.utexas.edu}

\author{Hanlin Li}
\authornotemark[1]
\orcid{0000-0002-0688-9918}
\affiliation{%
  \institution{University of Texas at Austin, School of Information}
  \city{Austin}
  \state{Texas}
  \country{USA}
}
\email{lihanlin@utexas.edu}

\renewcommand{\shortauthors}{Lee et al.}

\begin{abstract}
    Existing work on AI openness has focused on defining what technical components or release practices qualify a system as ``open''. However, less is known about how openness is understood and put into practice by people who adopt and adapt these models under real-world constraints. In this paper, we present an empirical study of r/LocalLLaMA, a large online community centered on running and customizing open foundation models locally. Through thematic analysis of community discussions, we find that members conceptualize openness pragmatically—in relation to reliability, local control, privacy, and the ability to adapt models under constraints such as compute resources, licensing, and usability. We identify key motivations for adopting open models—including autonomy, experimentation, and resistance to platform instability—as well as deterrents such as steep learning curves and performance gaps compared to closed systems. We further describe how shared resources and projects—including datasets, evaluation frameworks, and inference tools—sustain interdependent development in the broader open AI ecosystem beyond individual model releases. We then discuss the implications of a utility-oriented view of openness, and how producer support for downstream usability and infrastructure could better enable sustained innovation in open model ecosystems.
\end{abstract}




\maketitle

\section{Introduction}

Openness in AI is generally defined as the degree to which an AI system's components—including but not limited to training data, training code, and model parameters—are made available \cite{bostrom2018strategic, white_2024, liesenfeld_2023}. Researchers have advocated for open models because of their customizability, utility for research, and economic impact \cite{eiras_2024c}. Furthermore, open source models have enabled researchers to audit training datasets \cite{bolukbasi_2016}, develop fine-tuning methods that improve model safety \cite{bianchi_2024}, and discover brittleness in model safeguards \cite{arditi_2024a}. Beyond safety, open models have also driven general research innovation, spurring advances in the speed of inference \cite{kwon_2023}, efficient fine-tuning techniques \cite{rafailov_2024}, and high-quality human preference data \cite{kopf_2023a}.

While the benefits of open models for scientific research are well documented, less is known about how open models are received and utilized in the wild. In existing work, model repositories like Hugging Face and GitHub have been studied both as a marketplace and a community to understand collaborative development \cite{linakerCartographyOpenCollaboration2025a, castanoAnalyzingEvolutionMaintenance2024}, patterns of user behavior \cite{Choksi_2025}, and the economic value of open models \cite{longpreEconomiesOpenIntelligence2025}. However, prior work provides limited insight about how end user innovators participate in, interpret, and sustain these communities beyond formal development and model production.

To understand perspectives and practices around open models in the wild, we studied r/LocalLLaMA, an online community focused on locally running and adapting open models. This subreddit has been identified by Choksi et al. as a potential venue to understand user perceptions, practices, and collaboration around models \cite{Choksi_2025}. Through a mixed-methods analysis combining topic modeling and thematic coding, we analyze conversations surrounding open AI systems on the subreddit. To guide our inquiry, we ask: \textbf{RQ1}) How does r/LocalLLaMA conceptualize openness in AI?; \textbf{RQ2}) What are the drives and deterrents to adoption of open AI? ; and \textbf{RQ3}) How does the community collectively solve problems that arise in the use of open AI systems?

We find that members of r/LocalLLaMA articulate multiple, and conflicting, conceptions of openness. These definitions are tightly coupled to practical constraints---such as compute, licensing, and usability---and directly shape both adoption decisions and patterns of collective development. While ``open source purists'' demand strict replicability and specific licensing standards, ``pragmatists'' value any release that provides tangible utility, such as open weights, often prioritizing the ``spirit'' of openness over definitional rigor. Beyond these debates, we identify key drivers for adopting open models, including reliability, privacy through local inference, the ability to bypass prescriptive censorship, and opportunities for hands-on experimentation. Conversely, we report significant barriers to adoption, specifically the sharp learning curve associated with fragmented open-source software tooling and the performance advantages of proprietary systems in areas like multilinguality and resource efficiency.

Our findings suggest that prevailing definitions of openness---largely articulated from a production-centric perspective---do not fully capture how openness is enacted by those who adopt models in practice. For the r/LocalLLaMA community, openness is not a fixed set of released components but a dynamic property defined by what a model enables them to do under real-world constraints. This practical openness is realized through a process of assembly, where models are combined with community-led infrastructure. By shifting the focus from model artifacts to the cumulative innovation of the ecosystem, we show how users navigate resource asymmetries to maintain local control and reliability. Furthermore, we recommend model producers to lower the burden of this assembly layer by ensuring day-one compatibility with community tooling, allowing the community to shift its limited resources from remedial maintenance to generative innovation.

Throughout the paper, we use \textbf{foundation model} as a term to mean "any model that is trained on broad data (generally using self-supervision at scale) that can be adapted (e.g., fine-tuned) to a wide range of downstream tasks" \cite{bommasaniOpportunitiesRisksFoundation2022}
. We use the term \textbf{open model} as an inclusive term for open-weight foundation models and open-weight foundation models with their training data and code released openly. We also use the term \textbf{open source machine learning software}, or \textbf{MLOSS} for short, to denote infrastructural software that support both the training of as well as the inference, application, and deployment of models \cite{Langenkamp_Yue_2022}. 

\section{Related Work}

In this section, we first describe how definitions of openness in AI have broadened from its initial grounding in open source principles, to open science, open governance, democratization, and multidisciplinary perspectives. Then, we outline the known benefits and drawbacks of open AI systems. Lastly, we review current research on the community and developer dynamics of machine learning and AI related repositories, like GitHub and Hugging Face.

\subsection{Definitions of Openness in AI}

The term ``open source'' stems from Free Software principles \cite{fsf25} but has been adapted for foundation models, which contain non-software components like model weights and training data. Eleuther AI first applied this terminology to their GPT-NeoX-20B release, which included training code, data, and weights \cite{gaoPile800GBDataset2020, blackGPTNeoX20BOpenSourceAutoregressive2022}. As open efforts in building foundation models ensued, researchers began investigating what the term ``open source'' means for them, and evaluating them against a systematized definition. Liesenfeld et al. developed a set of 13 components of a model related to its openness, such as architectural documentation, data sheet, and open licensing \cite{liesenfeld_2024}. Others expanded the definition of model openness beyond open source. For example, White et al. incorporated ``open science''—the practice of making all stages of scientific research openly accessible, reusable, and transparent to anyone—and ``open tooling''—the full scope of libraries and tools needed to train, assess, and test models—into a tiered Model Openness Framework \cite{white_2024}. Furthermore, Paris et al. argued that other academic fields consider not only the properties and affordances of openness but also its \textit{intended effects} \cite{Paris_2025}. They advocated that evaluating intended effects will guide the creation of concrete and measurable goals—such as adding model cards to improve transparency or creating incentives for auditing to strengthen accountability \cite{Paris_2025}.

Existing literature remains primarily taxonomic, focusing on what technical components or institutional goals are required to satisfy formal definitions of openness. However, these frameworks overlook the realities of the users who actually adopt and adapt these models. We seek to address this gap by investigating the bottom-up conceptualizations of openness within a practitioner community, as well as external factors that they perceive to affect the degree of openness possible.

\subsection{Benefits and Drawbacks of Open AI}

The open AI ecosystem has expanded rapidly, from early replication efforts like GPT-NeoX \cite{black2022gpt} and collaborative initiatives like BLOOM \cite{workshop2022bloom} to industry-defining releases like LLaMA \cite{touvron2023llama} and fully open systems like OLMo \cite{olmo20242, groeneveld2024olmo}. Advocates argue that these models catalyze scientific progress, enabling critical external oversight and decentralizing control over a transformative technology \cite{eiras_2024c}. These open models enable researchers outside of industry labs to perform safety and interpretability research, where white-box access to model weights allow researchers to perform stronger attacks, and interpret models more thoroughly \cite{Casper2024BlackBoxAI}. Furthermore, open models are used to reproduce advancements in specialized applications like reasoning \cite{hu2025open}, and coding \cite{huang2024opencoder}.

However, the authenticity of openness in the ecosystem is frequently contested, with Liesenfeld et al. identifying a trend of "open-washing"---the practice of branding systems as open-source despite a lack of genuine transparency or public benefit \cite{liesenfeld_2024}. This skepticism extends to the broader narrative of democratization in openness; Seger et al. noted that most releases fail to democratize critical aspects like profits and governance \cite{seger_2023}, while Widder et al. argued that even ``maximally open'' releases fail to democratize R\&D because they still require datacenter-scale compute and privileged hardware access \cite{widder_2023}. As Howison et al. observed, successful coordination around open-source artifacts requires that they be cheap to instantiate and distribute \cite{HOWISON_2014}---conditions that remain largely unmet by contemporary foundation-model releases.

Alongside these structural critiques of ``open'' models, researchers highlight the irreversible risk associated with their weights being open. Unlike closed systems, open models lack an ``undo button,'' meaning their release can permanently enable malicious activities such as misinformation or large-scale scamming \cite{segerOpenSourcingHighlyCapable2023}. Furthermore, the ability to directly manipulate model weights allows users to bypass safety training entirely, modifying models so that they no longer refuse any request \cite{arditi_2024a}. A gap remains in this scholarship, which primarily evaluates openness through macro-level structural critiques or assessments of potential and theoretical risk. There is little empirical research into the specific motivations and deterrents that shape how users choose to adopt open models. To address this gap, we want to find how these known benefits as well as limitations manifest in the lived experiences of users, understand how they negotiate with them, and surface any so far unknown.

\subsection{Open AI Repositories and Communities}

Researchers have explored model repositories as a marketplace of models but also as a community. Langenkamp and Yue found that open sourcing machine learning (ML) related software libraries—like PyTorch and Transformers—has an effect of community creation, from which their developers collect feedback to direct prioritization of features and make better user experiences \cite{Langenkamp_Yue_2022}. Researchers have also analyzed development patterns and user behaviors on Hugging Face—the leading repository for ML models and datasets—and found that models are only developed by a handful of developers \cite{osborne_2024}, with concentrated popularity among authors who often collaborate with one another \cite{castanoAnalyzingEvolutionMaintenance2024}. Choksi et al. find that users are nomadic in nature, jumping from model to model as the newest model beats its predecessor \cite{Choksi_2025}. They further point out that nomadic activity causes models to obsolesce quickly, and invited researchers to investigate cross-model communities outside of repositories like Hugging Face that may show how communities work with more durable parts of the open AI ecosystem \cite{Choksi_2025}. Linaker et al. conducted interviews with developers of multiple high-visibility LLMs on Hugging Face, and found that collaboration in these projects extend to datasets, benchmarks, open source frameworks, leaderboards, knowledge sharing, and compute partnerships \cite{linakerCartographyOpenCollaboration2025a}. Longpre et al. reported a layer of ``developer intermediaries'' who focus on creating derivative models that either make them more efficient or serve as ``artistic expression'' \cite{longpreEconomiesOpenIntelligence2025}. Both of these recent studies on model marketplaces show that the open AI ecosystem has shifted from a repository of isolated artifacts to a complex web of interdependent projects, where specialized roles emerge to bridge the gap between base models and practical applications.

% think about similarities/differences of our findings compared to linaker, engage deeply

Despite these insights into the structural evolution of the ecosystem, little is known about how end users participate in, and sustain these communities beyond formal development and model production. Prior work largely focuses on repositories, leaving the user-driven networks through which models are adopted, adapted, and socially interpreted over time underexplored. This omission is notable in light of von Hippel's account of open source software as a \textit{user innovation network} \cite{vonhippel_2005}—innovation can emerge without centralized producers when users have strong incentives to solve their own problems, voluntarily reveal their solutions, and can cheaply diffuse them to others. While these dynamics are well-established in open source software, their applicability to contemporary open AI ecosystems remains underexplored. Our work aims to address this gap by examining r/LocalLLAMA as a community of user innovators organized around adopting and adapting open models for practical use across different contexts, offering a complementary view to repository-centered studies of open AI ecosystems.

\section{Methods}

We analyzed 134 posts and 2,270 comments from r/LocalLLaMA using a mixed-methods approach that combines computational topic modeling with thematic analysis. We utilized topic modeling to structure the large volume of data and sample discussions relevant to our research questions, followed by a qualitative analysis to uncover community perspectives on openness, adoption, and development practices.

\subsection{Dataset Sourcing}
We use posts and comments from the subreddit r/LocalLLaMA, which is a unique hub where casual users, enthusiasts, and industry practitioners actively engage in the use and development of open AI. Although named after Meta's LLaMA model release, r/LocalLLaMA has since evolved into a general-purpose hub for the discussion, adoption, and adaptation of all locally runnable open-weight models. Reddit has been a popular research site as it offers a rich, publicly accessible record of community-driven discussions around emerging technologies \cite{Qi2023ExcitementsACA, Gamage2022AreDCA}. Choksi et al. specifically identified r/LocalLLaMA as a key venue for understanding cross-model activity \cite{Choksi_2025}, which reinforced the salience of this community. As of April 28th 2025, the subreddit's size ranked among the top 1\% in all of Reddit, and is trusted as a source for evaluating new model releases by industry leaders \cite{andrejkarpathy[@karpathy]_2023, _r}.

We collected posts and comments from r/LocalLLaMA spanning March 10, 2023 to January 30, 2025. This period captures pivotal moments in the open AI ecosystem, from the release of the first LLaMA model that popularized local inference of LLMs, and DeepSeek-R1 which meaningfully rivaled the performance of state-of-the-art models. Data was sourced using the Arctic Shift API\footnote{Arctic Shift API: https://github.com/ArthurHeitmann/arctic\_shift} and cross-validated against a Reddit data dump from Academic Torrents\footnote{Academic Torrents: https://academictorrents.com/} to ensure completeness. We removed bot activity, non-substantive content (e.g., emojis, one-word comments), and all deleted or removed items in both datasets to honor user privacy. This process resulted in a final corpus of 52,518 posts and 731,615 comments.

\subsection{Topic Modeling and Sampling Strategy}

% To navigate the large volume of data and identify relevant discussions, we employed topic modeling as a filtering and sampling mechanism. We used a BERTopic pipeline \cite{grootendorst_2022a} to cluster posts, iteratively adjusting parameters to ensure topic coherence. From 56 topics that the model generated, we selected a subset of 23 topics (documented in Appendix \ref{app:topic}) that were directly relevant to our research questions. We also sampled from the outlier category (Topic -1) after observing it contained multi-topical threads highly relevant to the study. 

To navigate the large volume of data and identify relevant discussions, we employed topic modeling as a filtering and sampling mechanism. This approach was appropriate for our dataset given its scale (52,518 posts; 731,615 comments), which precluded exhaustive manual review, and allowed us to surface topically coherent subsets of discussions for deeper qualitative analysis. We used a BERTopic pipeline \cite{grootendorst2022bertopic}, which combines transformer-based document embeddings with dimensionality reduction and density-based clustering. Specifically, we encoded posts using the \texttt{all\-MiniLM\-L12\-v2} sentence-transformers model \cite{reimers2019sentence}, reduced the resulting embeddings with UMAP \cite{mcinnes2018umap}, and clustered them using HDBSCAN \cite{mcinnes2017hdbscan}. We iteratively adjusted UMAP and HDBSCAN parameters (e.g., \texttt{n\_neighbors}, \texttt{min\_cluster\_size}) to ensure topic coherence and separation. From 56 topics generated, we selected 23 directly relevant to our research questions (documented in Appendix \ref{app:topic}). We also sampled from the outlier category (Topic -1)---designated to documents too dissimilar from the core document clusters to be assigned to a regular topic---after observing it contained multi-topical threads highly relevant to the study.

Following a modified version of Gauthier et al.'s "topic-guided thematic analysis" \cite{gauthier_2022} methodology, we filtered the top 50 posts by comment count within the 23 selected topics. We excluded posts regarding exclusively closed models, operating system troubleshooting, or specific consumer hardware to reduce noise. This process yielded 55 posts from six topics related to concepts of openness and adoption (RQ1 and RQ2), and 79 posts from 17 topics related to development practices (RQ3).

To characterize the sampled corpus, we report the following statistics. The final sample spans 124 posts with a mean of 119.46 comments per post (SD = 109.97, range = 9-555), totaling 14,813 comments. Comments averaged 263.7 characters in length, which suggests substantive engagement. The 5,647 unique users in our sample represented 9.2\% of the 61,408 unique commenters in the full dataset, with members contributing a mean of 2.62 comments (median = 1, 25th percentile = 1, 75th percentile = 2). Overall, the sample reflects a typical online discussion dynamic characterized by concentrated participation among a small core of contributors.

\subsection{Thematic Analysis}

% We utilized two distinct qualitative frameworks to address the different theoretical aims of our research questions.

% To explore community conceptualizations of openness (RQ1) and adoption drivers (RQ2), we employed Thematic Analysis \cite{braun_2006} , moving through standard phases of familiarization, iterative coding, and theme development. For the analysis of collective problem-solving (RQ3), we extended this framework to capture project evolution. Specifically, we mapped relationships between themes into a sequence—conditions, actions, and consequences—to illustrate temporal and causal mechanisms.

To explore community conceptualizations of openness (RQ1) and adoption drivers (RQ2), we employed Reflexive Thematic Analysis, moving through standard phases of familiarization, iterative coding, and theme development. The first author conducted all coding, maintaining a trail of codebook revisions across iterations. The coding was carried out by a single researcher, and the research team held regular meetings throughout the process to interrogate emerging codes and themes, discuss analytic decisions, and assess whether thematic saturation had been reached. Interpretive disagreements raised in these meetings were resolved through discussion among the team. For the analysis of collective problem-solving (RQ3), we extended this framework to capture project evolution. Specifically, we mapped relationships between themes into a sequence—conditions, actions, and consequences—to illustrate temporal and causal mechanisms.

We outline the data flow through topic modeling, sampling, and thematic analysis in appendix \ref{app:flow}, and share our codebook in appendix \ref{app:rq12}, and \ref{app:rq3}. In appendix \ref{app:tool}, we share some details of a web tool we developed for qualitative coding data that is nested in threads.

\section{Findings}

Below, we unpack how members of r/LocalLLaMA conceptualize openness and the factors that affect openness. We then describe the drivers and deterrents for adopting open models. Finally, we show how members drive innovation through interdependence at multiple layers of the open AI ecosystem.

\subsection{How the Community Conceptualizes Openness}

Members of r/LocalLLaMA invoke several different mental models and definitions to make sense of what ``openness'' means in AI. Some adopt the strict criteria of open source as defined in software licensing—emphasizing full replicability and freedom to use, modify, and distribute. Others use openness more loosely to describe any system that affords meaningful access, transparency, or modifiability, even if it falls short of formal open source standards. This range of interpretations reveals tensions over what openness means in theory versus what it enables in practice.

\subsubsection{Applying ``Open Source'' to AI}\label{os_pure}

We observed two approaches to translate the established ``open source'' term in open source software to the AI setting—those who equate open source to replicability and demand full openness across model components, and those who scrutinize the former's shoehorning of open source into AI components. Because both camps seek to preserve the integrity of the term ``open source'', we call them open source purists. 

The first camp of purists emphasize the ability to re-build the software using the source material. Since every component needed for building the model must be present, they define open source AI as models that come with the code and datasets used for its training, as well as its weights. The ability to build from scratch is closely linked to the academic ideal of replicability—a standard in research where a model’s results can be reproduced independently using the original training data and code. One member, who is more familiar with this standard as a researcher, expressed their frustration about ``not a single paper since transformers'' being replicable.

% % [os/tr/open_source_replicability] id=t1_m9kk04z
% \begin{quote}
%     ''As a researcher I am sick of the use [of] the saying open source. You are not OS unless you are completely replicable. Not a single paper since transformers has been replicable.``
% \end{quote}

% The second purist perspective extends the logic of software definitions into AI by centering licensing as the decisive marker of openness. Just as the freedoms guaranteed by software licenses historically defined what counted as open source, some members argued that AI models should be judged by the same legal criteria, even if the objects in question—datasets, weights, and training pipelines—do not map neatly onto software artifacts.

The second purist perspective centers on the definitions codified in OSS licenses, applying them strictly to AI components. Proponents of this view note that traditional OSS licenses have never required that all assets---such as icons or other static resources---be open, only that users be free to substitute their own. Applying this logic to AI, they categorize model weights as pre-compiled artifacts that users are expected to bring themselves, rather than components that must be openly provided.

% [os/tr/open_through_license] id=t1_lsryg7q
% \begin{quote}
%     ''Open source doesn't place restrictions on how you can use the software. Meta's license does...a key part of being open source is granting the freedom to `use, study, change, and 
% distribute' however the end user wants.``
% \end{quote}

From this view, access to the inference code alone is enough to qualify an AI system as open source. As one member explained, \textit{``for no traditional open source definition that I'm aware of, it is a requirement that these assets also can be re-built by the user, only that the user may bring their own.''} This reading draws on the Open Source Definition's silence on non-code assets rather than any explicit exclusion, a gap this member interprets as permission to treat weights similarly.
% However, the second camp of open source purists questioned the shoehorning of open source into model components, arguing that model weights should be seen as ``hardcoded values,'' much like visual assets such as icons in OSS. From this view, access to the inference code alone is enough to qualify an AI system as open source. Since open source definitions for AI often lean on analogies to elements of traditional OSS, questioning the validity of those links can expose the definition's fragility when applied to AI models.

% % [os/tr/inference_code_source] id=t1_lsqchpq
% \begin{quote}
%     ``...the model is not the actual application. The source code of the application IS available under an open source license and CAN be modified and built by the user. From a software point of view, the model weights are an asset like a graphic or 3D geometry. For no traditional open source definition that I'm aware of, it is a requirement that these assets also can be re-built by the user, only that the user may bring their own.''
% \end{quote}

\subsubsection{Open Source Pragmatism}\label{sec:osprag}

In contrast to the purists who debate the adaptability of ``open source'' from software to AI, some adopted a more pragmatic lens—valuing what is shared, however limited, and focusing more on the tangible utility partial openness can enable. By orienting their conceptualizations of openness around practicality, these members constructed their own understanding of openness, and adopted a critical eye towards prescriptive definitions.

% [os/op/weights_more_valuable] id=t1_lsposxa
% \begin{quote}
%     ``Even if Meta release the whole dataset + code, its not like everyone in their bedroom can suddenly download + modify + run it...
%     If Meta want to call their stuff ''Open Source`` I don't really care, they are certainly currently greatly contributing to the OSS community. Releasing the full foundation model is in the spirit of ''Open Source`` in my personal opinion.''
% \end{quote}

This line of reasoning shifts attention away from strict definitional debates, and towards the economic value an open release provides. Members who adopted this stance focus less on the adherence to the traditional “open source” criteria and more on the pragmatic assessment of the artifact itself—what form of it delivers the most usable value to the community. They acknowledged that modifying the source requires significant investment of time and resource, and concluded little value for sharing the source code of LLMs. In one instance, a community member described the source of LLMs as ``economically useless'' and saw the resulting weights as far more valuable because they offer an accessible way for others to appropriate a model for their own purposes. Similarly, another member acknowledged the impracticality of retraining with limited compute, and appreciated Meta’s foundation model releases as upholding the “spirit” of open source.

Members also noted that the restrictive licenses---often criticized in partially open models---are permissive for most users in practice, since their commercial limits far exceed the value an individual—or even a typical business—is likely to generate from using the model. From this perspective, the openness that really matters to an individual is the extent of freedom it grants to them, and not whether the license falls under the definitions of open source AI.

More broadly, the pragmatic members worried that dismissing the already vibrant field of partially open models simply because of definitional rigor may stifle their development. In criticism of excessive pedantry, they argued, \textit{``it's not like Meta owes it to anyone, and without their llama [models] we wouldn't have half as many interesting [projects] as we have now... Can't we just all be at least a bit more grateful for what we've been offered for free so far?''}

\subsubsection{External Factors that Affect the Degree of Openness Possible}\label{ext-fac}
While the r/LocalLLaMA community's internal debates define how openness is understood, external forces also shape members' conceptualizations of openness. 

\textbf{Regulation} is seen as a dampening force to openness by imposing technical and legal requirements—such as SB-1047’s "kill switch" mandates—that are functionally incompatible with open-source distribution. Critics viewed ``safety'' as a strategic front for regulatory capture, aiming to monopolize data access and enforce centralized control. Some also viewed the open sourcing of models as an anticipatory strategy to render regulatory interventions impossible, arguing that by releasing weights, companies are ``preemptively making any legislation against AI a lot harder'' since ``enforcement becomes pointless'' once models are out in the wild.

% % [os/reg/regulation_as_strategy] id=t1_l3f372g
% \begin{quote}
%     ``By releasing weights as open source of very good models, [Meta] are preemptively making any legislation against AI a lot harder... Sure they can still make laws targeting the companies that train AI, since you still need very powerful hardware, but still, a lot of it is already out in the wild. The legislation becomes a lot more pointless.''
% \end{quote}

% Members questioned who should be the appropriate subject of regulation: AI developers, or bad actors that use them maliciously. They often drew parallels to other tools like books, code, or physical objects to challenge the idea that models themselves should be subject to constraints on their content or capabilities.

Members in our sample also endorsed the ability to fine-tune models freely, having a low perception of risk towards their misuse. While academic literature on AI safety considers this rapid, irreversible model diffusion a risk of open release, this view does not resonate with the community. To them, the bad-actor problem that regulations seek to solve is not tractable, because there will always be bad actors, regardless of model availability.

Members also pointed out potential risks in regulating the availability of open models. When models are made unavailable through trusted routes like Hugging Face and GitHub, they hypothesized the emergence of a black market with less secure channels for distribution, ironically increasing the risks that regulation meant to prevent. In their words, \textit{``It's one thing to trust code and data from trusted sources like HF and GH. It's another to trust things from a random torrent.''}

The \textbf{resource gap} between corporate model developers and users also affects perceptions of openness. LLMs are notoriously expensive to train, which leaves only a small number of large tech companies as major players. As one member put it, \textit{``Creating a good AI model...[is] no longer something that can be done in a garage or a small university lab, unlike how individuals could once improve classifiers for cats and dogs. I'm not sure how open source with little funding could compete.''}

The number of users that a closed platform can access to is another resource gap, which provides valuable data for learning people's preferences. For example, one member noted that the user data OpenAI collects from their ChatGPT service is unparalleled to companies building open models, calling for a \textit{``renaissance in feedback loops and real world usage''} in the open AI community.

Because resources are scarce, members of r/LocalLLaMA remained attentive to what \textbf{corporate motives} exist for releasing models openly. For one, they acknowledged that commoditizing the complement, i.e. making software free and capitalizing its complements, to be a tactic to make open source efforts profitable. They also acknowledged that Meta gained an early-mover advantage in releasing its Llama model publicly, as it catalyzed a wave of derivative projects which still continue to be industry standards.

% [prof/osbs/open_source_profit] id=t1_l3d0wvx
% \begin{quote}
%     ``...llama.cpp, Vicuna, WizardLM, Unsloth, ExLlama, and a myriad of other projects now exist which Meta didn't have to create themselves. It's not (just) out of the goodness of their hearts, they are getting a financial benefit and are actively saving money by letting their competitors use their work.''
% \end{quote}

% Going beyond immediate effects on the open AI ecosystem, members noted the corporate incentive of talent acquisition. By releasing models publicly, corporations can showcase their technical leadership and generate enthusiasm among developers, ensuring that the best researchers arrive pre-trained on their tools and culture. In a market where talent scarcity drives salaries into the millions, one member noted that the perception of being the ``cool kids on the block'' carries a strategic weight.

Notably, members in our sample did not see these corporate motives as inherently malicious. Instead, they argued that so long as corporations are invested in open-source ecosystems, the arrangement still embodies the reciprocal ethos of open source. What matters to them is not purity of intent, but the maintenance of a mutually beneficial cycle. In fact, many members directly participated in the marketplace enabled by open models, offering custom solutions, fine-tuned systems, or on-premise deployments for businesses that seek practical value from less restrictive licenses. In this way, openness directly translates into entrepreneurial benefits for open model users.

\subsection{Drives and Deterrents to Adoption}\label{dnd}

While definitions of openness vary, the community's adoption of open AI is driven by a desire for control, privacy, and unrestricted utility. However, these benefits are frequently weighed against significant barriers in usability and performance.

\subsubsection{Data Sovereignty: Reliability, Privacy, and Offline Access}\label{sovereign} Data and compute sovereignty---the ability to control the model's environment and data flow---was a recurring theme for adoption across our posts. Members valued the consistency of local systems over proprietary APIs, which are perceived as opaque and unreliable. One member noted that unannounced API updates broke their workflows, while another experienced wildly non-deterministic outputs from the same API given an identical prompt.

This desire for control extends to privacy. As one member put it, local inference eliminates the need to send sensitive code or personal data to \textit{``some American company that's getting rich off my data.''} This is particularly critical in sensitive domains; one member described their application of LLMs within therapy that required local inference to protect patient privacy.

\subsubsection{Low Marginal Costs Enable Experimentation}\label{low-marginal} The one-time hardware investment for local inference encourages experimentation by removing the burden of recurring fees or metered API credits. As one user put it, \textit{``it's cheaper to play/research with smaller models... than renting computing power.''} This low marginal cost transformed open models into tools for experimentation, allowing users to \textit{``open a rabbit hole''} into machine learning and building complex agentic workflows without fear of accumulating costs. Users also noted efficiency in batch processing; one member calculated that processing 8B tokens locally cost them \$20 in electricity versus \$400 via the cheapest API provider.

\subsubsection{Uncensored Models and Ideological Freedom}\label{uncensored} Members in our sample criticized the ``paternalistic'' safety filters of proprietary models, arguing that limiting the capabilities of model outputs for the sake of safety is like \textit{``a dull knife that can't cut.''} To these critics, open-weight models presented a welcome alternative that allowed them to define their own safety standards. This drives the popularity of ``uncensored'' fine-tunes, which are used for applications ranging from horror story writing to Erotic Role-Play (ERP). Members defended ERP as a functional outlet and a safer alternative to conventional pornography.

\subsubsection{The Performance and Usability Gap}\label{pugap} Despite these motivations, members also expressed significant barriers to adoption.

First, members in our sample frequently encountered significant usability friction. The ``sharp learning curve'' of the open ecosystem---characterized by fragmented GitHub repositories and complex installation steps---contrasts poorly with the seamless experience of proprietary platforms like ChatGPT. One user summarized this frustration regarding the installation of open models: \textit{``I have a copy of firefox running on my private hardware. Why is it any harder to do it with LLM? ... Honestly, I don't even need to go further than this to know it is a usability nightmare.''}

Second, members identified capability deficits in open models compared to proprietary ones. While open models can outperform proprietary systems in niche tasks, members lamented that they generally lagged in general capabilities. More specifically, some cited a lack of robust multilinguality and multicultural knowledge as a deterrent, noting that proprietary models process diverse languages more effectively.

Finally, the choice of model is often dictated by resource asymmetry. The efficiency of centralized compute often outweighs local setups for heavy workloads. Members recognized that for massive context windows, proprietary APIs remain more economical due to economies of scale. Another noted that covering the cost of consumer hardware (Apple's M4 chips) would take \textit{``6 years of running... at its absolute limit''} compared to using OpenAI's models. Thus, while open models offer sovereignty, proprietary systems remained the tools of choice for efficiency.

\subsection{Adaptation of Open Models}\label{adaptation}
Open models in the r/LocalLLaMA community are valued primarily as tools to achieve practical applications. Unlike proprietary systems that often provide an isolated and vertically-integrated product, applications of open models are built through layered interdependence between people who train models, develop inference tools, and evaluate performance. Innovation occurs cumulatively through loosely coordinated roles, where partial contributions spark community interaction.

This interdependent dynamic can be understood through three lenses: the \textbf{condition} in the form of partial contributions that give rise to community engagement, the \textbf{actions} by which community members respond to these contributions, and the \textbf{consequences} that emerge as contributions accumulate.

\subsubsection{Conditions: Partial Contributions as Starting Points for Discussion}\label{condition}

Many posts on r/LocalLLaMA share partial or preliminary contributions that spark further community involvement. These contributions often come in the form of models, tools, or data that are useful but incomplete, implicitly inviting others to build on them. For example, we observed cases of:

\begin{itemize}
    \item \textbf{Fine-tuned models released without broader support}: One member shared a fine-tuned language model named \texttt{WizardVicunaLM} along with their novel training method, but provided no scaled-up versions or smaller quantized models for wider use, which left a \textit{scaling gap}.
    \item \textbf{Datasets without direct applications}: Another contributor released a distilled training dataset intended to compress a model's knowledge, yet did not publish any resulting smaller model or usage example, leaving a \textit{gap in application}.
    \item \textbf{Quantized models lacking evaluation}: Community developers produced optimized low-precision model files to enable local usage, but often without standardized quality evaluations. This created an \textit{evaluation gap} since users could not be sure how these 4-bit or 8-bit models performed relative to originals.
    \item \textbf{Tools introducing compatibility issues}: Developers shared new versions of inference engines that improved performance but broke compatibility with older model formats or pipelines. Each such update highlighted a \textit{compatibility gap} for users who then struggled to keep their local setups working.
    \item \textbf{Applications without open-source code}: Some posts demonstrated impressive local AI applications, like a personal voice assistant or a mobile chatbot, yet initially withheld the source code. This limited the community's ability to extend or trust these apps, revealing an \textit{openness gap}.
    \item \textbf{RAG systems without evaluation standards}: Similar to above, a few members showcased retrieval-augmented generation tools but provided no clear way to evaluate their accuracy or reliability. This absence of metrics created an \textit{evaluation gap} for RAG systems as well.
\end{itemize}

In each of these scenarios, a member shares an innovation that is valuable but partial, exposing a specific unmet need---whether it be scaling, thorough evaluation, or open access. These gaps become the starting points for community discussion and drive the next phase of interaction.

\subsubsection{Actions: How the Community Responds to Identified Gaps}

When such gaps are identified, the r/LocalLLaMA community tends to respond in patterned ways. Members critically discuss limitations, propose improvements or standards, and encourage those who make useful contributions. These condition $\rightarrow$ action responses repeat across different posts and project types.

First, community members \textbf{openly articulate the limitations} or risks of the partial contributions. This often takes the form of critical feedback or requests in the comment threads. For instance, after the \texttt{WizardVicunaLM} model was shared, several commenters immediately asked for its training method to be applied to other model sizes (e.g. 7B or 13B parameters), implicitly pointing out the \textit{limited scale} of the original release. In other cases, when someone posted impressive benchmark results for a model, experienced users called out possible cherry-picking of tasks or noted where the model still failed---highlighting hidden weaknesses behind the reported averages. There were also reports of highly-touted ``benchmark'' models failing specific user tests, which users did not hesitate to share, tempering any unrealistic expectation. When a new version of an inference tool introduced breaking changes, many voiced frustration and criticized the lack of backward compatibility, directly asking maintainers to address this stability issue. Likewise, upon seeing a RAG demo without any evaluation, commenters pointedly noted the absence of proper benchmarks to measure its effectiveness. Beyond individual cases, some discussions took a step back to point out broader neglect in the ecosystem---for example, arguing that the community was focusing too much on training ever-larger models while tooling, interfaces, and evaluation methodologies lagged behind. By candidly identifying these issues, the community creates a shared understanding of what is missing or could go wrong.

The next common action is \textbf{proposing solutions or improvements} to address the gaps. Community members rarely stop at just pointing out problems; they brainstorm and suggest concrete steps forward. Often these suggestions extend or modify the original contribution rather than replacing it entirely. For example, to tackle the recurring compatibility problems with model files and loaders, users collectively endorsed the adoption of the new GGUF model format (a successor to GGML) as a long-term stability solution. This push for a standardized format was a direct response to the breakdowns caused by frequent format changes. In another instance, after recognizing the lack of evaluation for RAG systems, a group of commenters engaged in ideation on how to evaluate RAG consistently---discussing possible shared datasets or comparison protocols that creators could use in future posts. We also saw specific feature suggestions: in a post asking for a wishlist of features in a future release of Google's Gemma model, users suggested improvements such as adding multilingual support, user-defined personas, and relaxing overly strict content filters. These ideas aimed to make the tool more flexible and globally usable, addressing its perceived shortcomings. It is important to note that these community-suggested fixes or extensions are usually incremental. They seldom provide a complete or immediate remedy to the gap; instead, they serve as directional nudges. In practice, such proposals inform subsequent updates or inspire new projects, gradually steering the community’s collective work toward more robust solutions.

The community’s third type of action is \textbf{positive reinforcement} for those who address gaps or contribute useful building blocks. Recognition and encouragement play a key role in sustaining open-source efforts. For instance, when one member shared a curated reasoning dataset a large model's outputs or a comprehensive compilation of evaluation results across models, many replies expressed gratitude and emphasized how valuable that work was. Posts introducing helpful tools or applications also garnered significant praise: the developer of \texttt{PocketPal} – a mobile app enabling on-device LLM chat---received enthusiastic feedback for making local AI more accessible, and users thanked them especially once the app was open-sourced. Similarly, an individual who built an open-source clone of the Perplexity search engine (allowing users to query the web with a local LLM) was applauded for the initiative. This kind of community response reinforces the importance of filling the identified gaps. It signals to contributors that efforts spent on often less glamorous tasks like optimization, evaluation, and integration are highly appreciated. Such encouragement can motivate creators to continue improving their projects or inspire others to embark on similar contributions, knowing there is support and demand.

\subsubsection{Consequences: Interdependent Cooperation}\label{consequence}

Over time, the cycle of partial contributions and community responses leads to cumulative, interdependent outcomes. What begins as one person’s project can evolve into collaborative endeavors, and new supporting roles or artifacts emerge. These are the \textit{consequences} of the repeated condition $\rightarrow$ action loops, gradually knitting individual efforts into a richer ecosystem.

One major outcome is that promising ideas get \textbf{scaled up through community involvement}. A clear example is the \texttt{WizardVicunaLM} case: the original author lacked resources to train the model at larger scales, but after interest grew, several other members donated their compute power to train 7B, 13B, and even 30B-parameter versions of the model. They then shared these larger models in various quantized formats back with everyone. In effect, what started as a single fine-tuning experiment became a family of models produced by multiple volunteers, dramatically amplifying its impact. We saw a similar scaling effect in the quantization domain. When one person investigated the performance of different quantization techniques on a model, others joined in by contributing additional test data and results. The community collectively added data points to benchmark 8-bit, 6-bit, 5-bit, and 4-bit quantized models, improving confidence in the findings. This pooling of efforts meant no single contributor had to do everything; initial gaps like the need for larger models or more tests can be filled through distributed contributions.

Another consequence is that projects often \textbf{evolve beyond their initial scope} based on community feedback. For instance, the creator of the Perplexity clone returned to their project after launch to add features and fix issues highlighted by users, making the tool more robust and useful than the initial version. In cases where a contributor’s claims or results were challenged---say, a model’s benchmark was criticized for missing certain tests---the original poster usually responded by refining their work, such as running new evaluations or releasing improved model versions to address the community’s concerns. Community enthusiasm can even change a project’s accessibility: developers of a local voice assistant and the PocketPal app both decided to open-source their code after seeing strong interest and encouragement from commenters. What might have remained personal or closed projects thus transformed into public resources that others could modify or build upon. In the long run, some of these tools originating from single users have grown into shared infrastructure, maintained by multiple community members or widely adopted as a foundation for new experiments.

Finally, the repeated interactions give rise to \textbf{new specialized roles and spin-off projects} that aim to permanently address recurring gaps. A notable example is how the community handled the ongoing instability in inference tooling. Instead of relying solely on the original maintainers, one experienced member forked the \texttt{llama.cpp} repository into a variant called \texttt{KoboldCpp} to focus on a more stable client. Similarly, the collective brainstorming about RAG evaluation didn’t end at the comment-level: it directly inspired a contributor to create a new \texttt{AutoRAG} toolkit designed to automate retrieval-augmented generation workflows and help standardize their evaluation. In essence, community discussions crystallized into a new artifact that filled the evaluation gap for RAG. These developments show how, over time, r/LocalLLaMA generates its own ecosystem support structures. Enthusiastic individuals step up as maintainers, tool builders, or coordinators in specific niche areas---be it quantization techniques, benchmarking, or interface stability---thereby creating solutions to problems that were initially raised informally.

Notably, this vibrant ecosystem growth happened without any formal coordination or top-down planning. Instead, it arose through transparent sharing of work and open dialogue. When partial artifacts are posted and their limitations openly discussed, it enables anyone with interest or resources to jump in and contribute. Over time, the community’s iterative contributions knit together: models feed into applications, tools enable new experiments, and evaluations inform improvements, creating a layered, interdependent network of open AI development.

\section{Discussion} \

This section situates our findings within the broader open AI discourse, arguing that users prioritize pragmatic utility and sovereignty over formal definitions of openness. We show how this drive for autonomy fosters durable cross-model infrastructure and propose a new division of labor to better sustain distributed innovation.

\subsection{Defining Openness Through Pragmatic Utility}
% Our findings show that users of r/LocalLLaMA are acutely aware of the limits of what is commonly labeled ``open'' in contemporary foundation models. While some members adhered to strict open-source definitions---emphasizing replicability, licensing, and access to training data and code (Section 4.1.1)---many explicitly acknowledged that retraining large models from scratch is infeasible under realistic compute constraints (Section 4.1.2). As a result, openness is not evaluated by whether all components are released, but by whether the release enables meaningful use.

% Rather than being misled by ``open-washing,'' \cite{Liesenfeld_2024} community members often looked past formal definitions and focused on the practical value of released artifacts. We observed repeated claims that model weights are the most economically useful component, while full training code or datasets offer little marginal benefit to individual users (Section 4.1.2). Even restrictive licenses are frequently treated as irrelevant in practice, as their limits exceed what most users expect to derive from the model. In this sense, openness is understood as a \textit{means to act}, not a property to be certified.

Liesenfeld et al. have described ``open washing'' as the practice of marketing models as open-source despite lacking genuine transparency or public benefit, particularly in regulatory contexts such as the EU AI Act \cite{liesenfeld_2024}. 
Relatedly, others have argued that use restrictions and competitor exclusions in Meta's LLaMA licenses violate open-source standards and function as anti-competitive strategies \cite{tarkowskiMirageOpenSourceAI2023, _2023a, marisMetasLLaMaLicense2025}. These accounts characterize open washing as a deceptive practice that risks misleading stakeholders about a model's true degree of openness.

% These accounts suggest a form of deception: that stakeholders are at risk of being misled into believing such models are more open than they actually are.

Our findings show that members of r/LocalLLaMA are highly cognizant of the definitional shortcomings of many ``open'' model releases, explicitly debating missing components such as training data, code, and licensing constraints (Section \ref{os_pure}). Users routinely acknowledge that retraining foundation models from scratch is infeasible under real-world compute and data constraints, and that access to full training pipelines offers little practical value to most end users (Section \ref{sec:osprag}). As a result, openness is evaluated in terms of the concrete utility of released artifacts---most notably model weights---rather than adherence to formal criteria.

This does not diminish the importance of guarding against the misuse of the term ``open source,'' but it suggests the need for more precision in how open washing is applied. In this community, partial openness is neither obscured nor passively accepted; it is actively negotiated and often justified in light of structural constraints and practical goals (Section \ref{ext-fac}). The key implication is that openness functions less as a binary property and more as a condition shaped by what users can realistically do with a model. Recognizing this helps to explain why partially open releases can still meaningfully support downstream innovation, and why critiques focused solely on formal definitions risk overlooking how openness is enacted in practice.

% This pragmatic stance is reinforced by structural constraints external to the community. Members explicitly referenced the concentration of compute, data, and user feedback within large firms as an enduring barrier to full democratization (Section 4.1.3). Yet, this did not diminish the perceived value of open-weight releases. Instead, it reframed openness as something that is necessarily partial, but still sufficient to support downstream experimentation and reuse. This perspective underpinned the rest of the findings: once training from scratch was seen as out of scope, effort shifted toward building with models rather than rebuilding them.

\subsection{Data and Compute Sovereignty Under Openness Limitations}

Prior work has emphasized that openness does not equate to democratization. Seger et al. argued that open releases rarely distribute control over profits or governance \cite{seger_2023}, while Widder et al. showed that even maximally open models remain inaccessible without datacenter-scale compute \cite{widder_2023}. These critiques rightly point out that openness alone does not resolve systemic inequalities in AI development. What they leave underexplored, however, is how end users experience and act within these constraints---particularly what forms of value remain if full democratization is off the table.

Our findings show that despite limitations in openness, open models provide data and compute sovereignty: control over the model's execution environment, data flow, and operational stability (Section \ref{sovereign}). Members contrast local inference with proprietary APIs that are perceived as opaque, unstable, and privacy-invasive. Even when openness is partial, users value the ability to run models offline, prevent sensitive data from leaving their machines, and avoid silent updates that alter system behavior. This sovereignty is tied to practical outcomes such as privacy preservation in sensitive applications and predictable cost structures enabled by one-time hardware investments rather than recurring usage fees (Section \ref{low-marginal}). In this sense, open models function less as vehicles for democratization and more as tools for autonomy under constraint.

% The properties of autonomy our findings document — local data control, operational independence, and resistance to silent behavioral changes — do not map cleanly onto existing legislative frameworks; neither GDPR nor the EU AI Act directly addresses local model inference as a deployment paradigm, representing a gap in how current governance structures account for end-user autonomy in AI.

Our findings also complicate dominant safety framings in the literature. Previous work examining the risk potential of open models highlight the irreversibility of open releases and the risk that users can undo safety training or bypass safeguards \cite{segerOpenSourcingHighlyCapable2023, eiras_2024c, kapoor_2024a}. In contrast, we observe users who explicitly seek these properties and frame them as benefits rather than risks (Section \ref{uncensored}), grounded in a rejection of paternalistic control over acceptable use.

At the same time, we clearly identify barriers that limit adoption: steep learning curves, performance gaps relative to proprietary systems, and resource constraints that make local deployment impractical in some scenarios (Section \ref{pugap}). We conclude that for users, the adoption problem is a balancing act between these drives and deterrents: sovereignty and flexibility, against convenience and performance.

These findings also carry implications for policy and industry. Regulatory frameworks calibrated to deployment context rather than training scale alone would better preserve the sovereignty benefits our findings document, avoiding restrictions that eliminate local inference without meaningfully reducing risk. Model producers, similarly, could formally recognize these use patterns by offering stable, privacy-preserving local deployment options as a baseline release commitment.

\subsection{Developing Durable Tools and Infrastructure Across Model Generations}

Once adopted, open models are rarely used in isolation. Our findings show a dense ecosystem of community-built models, tools, and evaluation practices layered on top of base model releases (Section \ref{adaptation}). These models---whether instruction-tuned or not---serve as substrates for derivative fine-tunes, quantizations, inference backends, and applications. Innovation proceeds through partial contributions that expose gaps, which then attract responses from others who extend, stabilize, or generalize the original work. This directly answers calls by Choksi et al. to examine cross-model communities that persist beyond the rapid obsolescence of individual model releases \cite{Choksi_2025}.

Durable, cross-model artifacts were commonly praised by members. Techniques and tools that apply across multiple base models---such as inference engines (e.g. \texttt{llama.cpp} and its variants), standardized model formats, or RAG evaluation tools---are repeatedly treated as high-impact contributions (Section \ref{consequence}). These findings align with prior observations that open AI development extends beyond models to include benchmarks, tooling, and evaluation infrastructure \cite{linakerCartographyOpenCollaboration2025a}, while showing how such work is carried out by downstream users and developer intermediaries \cite{longpreEconomiesOpenIntelligence2025}. In contrast to repository-centered views that emphasize model churn and their concentrated authorship, r/LocalLLaMA members invest effort in infrastructure that remains useful across model generations. Our study empirically documents how these artifacts emerge, why they are built, and how they evolve through repeated condition-action-consequence cycles within the community.

The durability of these community-built artifacts also raises questions about their relationship to formal governance and industry practice. Tools like llama.cpp has already been adopted into commercial pipelines, suggesting that community infrastructure can transition from informal workaround to recognized dependency. Governance frameworks that account for this pathway—whether through artifact attribution standards, maintainer support programs, or inclusion in official model release documentation—would better reflect how open AI development actually functions in practice.

\subsection{Implications for Producer Support of Downstream Infrastructure}

These findings suggest a need to reconsider how labor is distributed between model producers and downstream communities. Currently, users expend substantial effort on remedial infrastructure---debugging inference issues, producing quantizations, ensuring backend compatibility, and constructing benchmarks (Sections \ref{pugap}, \ref{condition}). Much of this work compensates for gaps left at release.

Model producers could reduce this burden by investing more in downstream usability: ensuring day-one compatibility with widely used backends, releasing official benchmarks for quantized models, and stabilizing interfaces. While this may appear to conflict with the idea that openness leverages distributed innovation, our findings suggest the opposite. We posit that lowering the marginal cost of setup will enable the community to direct more of their efforts toward higher-order innovation.

We argue that producers who support this shift stand to benefit from a richer reciprocal relationship. Our findings show that when basic infrastructural gaps are identified and addressed through community action, effort shifts toward cumulative work, following the condition--action--consequence cycles described in Section \ref{adaptation}. In these cases, partial contributions become foundations for scaling and generalization: promising fine-tunes are expanded across model sizes through pooled compute (e.g., \texttt{WizardVicunaLM}), confidence in low-precision models is built through collective quantization testing, and recurring pain points give rise to durable tools via forks and new projects (e.g., \texttt{KoboldCpp} and \texttt{AutoRAG}).

% Seen this way, producer investment in downstream infrastructure would not replace user innovation, but alter the conditions under which these dynamics unfold. By reducing the recurrence of the same infrastructural gaps, producers could enable community effort to concentrate on the durable, cross-model contributions that our findings show to compound over time.

Seen this way, producer investment in downstream infrastructure would not replace user innovation, but alter the conditions under which these dynamics unfold. By reducing the recurrence of the same infrastructural gaps, producers could enable community effort to concentrate on the durable, cross-model contributions that our findings show to compound over time. Platforms such as Hugging Face are equally positioned to operationalize pragmatic utility as a release standard: surfacing release completeness metadata---backend compatibility, quantization readiness, available benchmarks---would reduce the investigative burden currently absorbed by communities like r/LocalLLaMA, where users routinely reconstruct this information through collective trial and error. Lowering the threshold at which pragmatic utility is legible to end users would enable communities to direct effort toward higher-order innovation rather than remedial infrastructure.

% We see evidence that when basic infrastructure is in place, users are able to move toward more generative work: scaling promising fine-tunes across model sizes through pooled compute (e.g. WizardVicunaLM expansions), improving confidence in low-precision deployments by collectively testing quantization methods, and turning divergent needs into durable tools through forks and new projects (e.g., KoboldCpp for stability and AutoRAG in response to evaluation gaps). In this way, we envision that producer investment in downstream infrastructure will not necessarily replace user innovation; rather, it will shift community effort toward contributions that compound across models and over time.

\subsection{Acknowledgements}

This research was partially supported by the following: the National Science Foundation DGE-2125858 grant; Good Systems, a UT Austin Grand Challenge for developing responsible AI technologies\footnote{\url{https://goodsystems.utexas.edu}}; and UT Austin’s School of Information.

\bibliography{references}
\bibliographystyle{ACM-Reference-Format}

\clearpage
\appendix

\section{Posts with resource links}
\begin{longtable}{p{6cm}lp{3cm}}
\caption{Posts with resource links} \label{tab:posts} \\
\toprule
Post title & Resource links & Project type \\
\midrule
\endfirsthead
\caption[]{Posts with resource links} \\
\toprule
Post title & Resource links & Project type \\
\midrule
\endhead
\midrule
\multicolumn{3}{r}{Continued on next page} \\
\midrule
\endfoot
\bottomrule
\endlastfoot
Open LLM Leaderboard new interface & \href{https://huggingface.co/spaces/open-llm-leaderboard/open\_llm\_leaderboard}{Hugging Face} & Review/leaderboard \\
UGI-Leaderboard remake & \href{https://huggingface.co/spaces/DontPlanToEnd/UGI-Leaderboard}{Hugging Face} & Review/leaderboard \\
SQL generation benchmark & \href{https://github.com/petavue/NL2SQL-Benchmark}{GitHub} & Review/eval \\
GPT-4 alternatives &  & Review, Inference \\
Open source coding assistants? &  & Review, Inference \\
Comparing different whisper packages & \href{https://amgadhasan.substack.com/p/sota-asr-tooling-long-form-transcription}{Blogpost} & Review \\
Experience with Codestral for Android development &  & Review \\
Huge LLM Comparison &  & Review \\
LLM "serious use" comparison &  & Review \\
Open-source Perplexity alternative & \href{https://github.com/shadowfax92/Fyin}{GitHub} & RAG \\
Yet another RAG system & \href{https://github.com/snexus/llm-search/tree/main}{GitHub} & RAG \\
How we chunk &  & RAG \\
llama.cpp web search integration & \href{https://github.com/TheBlewish/Web-LLM-Assistant-Llama-cpp}{GitHub} & RAG \\
AgentSearch & \href{https://github.com/SciPhi-AI/agent-search}{GitHub}, \href{https://huggingface.co/SciPhi/Sensei-7B-V1}{Hugging Face} & RAG \\
A common misconception about RAG frameworks &  & RAG \\
Collection of open source RAG techniques & \href{https://github.com/NirDiamant/RAG_Techniques}{GitHub} & RAG \\
AutoRAG & \href{https://github.com/Marker-Inc-Korea/AutoRAG}{GitHub} & RAG \\
EntityDB & \href{https://github.com/babycommando/entity-db}{GitHub} & RAG \\
RAGBuilder & \href{https://github.com/kruxai/ragbuilder}{GitHub} & RAG \\
Your best RAG projects &  & RAG \\
Perfect labels via prompting 2 &  & Prompting \\
Axiom prompt engineering & \href{https://github.com/codedidit/axiomprompting}{GitHub} & Prompting \\
Misguided attention & \href{https://github.com/cpldcpu/MisguidedAttention}{GitHub} & Prompting \\
Trailing whitespace in prompts &  & Prompting \\
Llama 2 prompt format & \href{https://huggingface.co/blog/llama2#how-to-prompt-llama-2}{Blogpost} & Prompting \\
Perfect labels via prompting &  & Prompting \\
Prompt format comparisons &  & Prompting \\
Open models wishlist &  & Model/wishlist \\
Local translator based on LLaMA & \href{ttps://github.com/OpenBuddy/OpenBuddy, https://github.com/dustinchen93/text-generation-webui-translator}{GitHub} & Model/translation \\
Finetune LLaMA2 for any language & \href{https://github.com/UnderstandLingBV/LLaMa2lang}{GitHub} & Model/translation \\
A paradigm shift in machine translation &  & Model/translation \\
Translate to 400+ languages & \href{https://huggingface.co/jbochi/madlad400-3b-mt, https://huggingface.co/spaces/jbochi/madlad400-3b-mt}{Hugging Face} & Model/translation \\
Quantizing Llama 3 8B seems more harmful &  & Model/quant \\
1.58bit Deepseek R1 & \href{https://huggingface.co/unsloth/DeepSeek-R1-GGUF}{Hugging Face}, \href{https://unsloth.ai/blog/deepseekr1-dynamic}{Blogpost} & Model/quant \\
exl2 quantization & \href{https://github.com/turboderp-org/exllamav2}{GitHub}, \href{https://huggingface.co/LoneStriker/Aetheria-L2-70B-2.4bpw-h6-exl2-2}{Hugging Face} & Model/quant \\
Unsloth dynamic quantization & \href{https://huggingface.co/unsloth/QwQ-32B-Preview-unsloth-bnb-4bit}{Hugging Face}, \href{https://colab.research.google.com/drive/1T5-zKWM_5OD21QHwXHiV9ixTRR7k3iB9?usp=sharing}{Notebook}, \href{https://unsloth.ai/blog/dynamic-4bit}{Blogpost} & Model/quant \\
Llama 3 quants comparison & \href{https://github.com/matt-c1/llama-3-quant-comparison}{GitHub} & Model/quant \\
Mistral-Large 35GB quantization & \href{https://huggingface.co/ChenMnZ/Mistral-Large-Instruct-2407-EfficientQAT-w2g64-GPTQ}{Hugging Face} & Model/quant \\
Qwen2.5 14B quant comparison &  & Model/quant \\
Qwen2.5 32B quant comparison &  & Model/quant \\
Benchmarking effects of quantization & \href{https://github.com/jd-3d/MPA_Bench/tree/main}{GitHub} & Model/quant \\
SQLCoder & \href{https://github.com/defog-ai/sql-eval}{GitHub}, \href{https://huggingface.co/defog/sqlcoder-34b-alpha}{Hugging Face} & Model/finetune, Review/eval \\
OpusV1 models for story-writing and role-playing & \href{https://huggingface.co/dreamgen/opus-v1.2-7b}{Hugging Face}, \href{https://colab.research.google.com/drive/1J178fH6IdQOXNi-Njgdacf5QgAxsdT20?authuser=1}{Blogpost} & Model/finetune, Model/dataset \\
LLM fine-tuning datasets & \href{https://github.com/mlabonne/llm-datasets}{GitHub} & Model/finetune \\
WizardVicunaLM & \href{https://github.com/melodysdreamj/WizardVicunaLM}{GitHub}, \href{https://huggingface.co/datasets/junelee/wizard_vicuna_70k, https://huggingface.co/junelee/wizard-vicuna-13b}{Hugging Face} & Model/finetune \\
llama-gaan-2-7b-chat-hf-dutch & \href{https://huggingface.co/Mirage-Studio/llama-gaan-2-7b-chat-hf-dutch}{Hugging Face} & Model/finetune \\
Medical LLMs for 50 languages & \href{https://github.com/FreedomIntelligence/ApolloMoE}{GitHub}, \href{https://huggingface.co/collections/FreedomIntelligence/apollomoe-and-apollo2-670ddebe3bb1ba1aebabbf2c}{Hugging Face} & Model/finetune \\
OpenBioLLM-70B and 8B & \href{https://huggingface.co/aaditya/OpenBioLLM-Llama3-70B , https://huggingface.co/aaditya/OpenBioLLM-Llama3-8B}{Hugging Face} & Model/finetune \\
WizardLM-13B-Uncensored & \href{https://huggingface.co/ehartford/WizardLM-13B-Uncensored}{Hugging Face} & Model/finetune \\
Lexi Llama-3-8B-Uncensored & \href{https://huggingface.co/Orenguteng/Lexi-Llama-3-8B-Uncensored}{Hugging Face} & Model/finetune \\
SmallThinker-3B-Preview & \href{https://huggingface.co/datasets/PowerInfer/QWQ-LONGCOT-500K, https://huggingface.co/PowerInfer/SmallThinker-3B-Preview}{Hugging Face} & Model/finetune \\
Unsloth & \href{https://github.com/unslothai/unsloth}{GitHub} & Model/finetune \\
Do you guys finetune models? &  & Model/finetune \\
How many people are fine tuning? &  & Model/finetune \\
Dataset of ~100k Japanese Web Novels and their English translations & \href{https://huggingface.co/datasets/NilanE/ParallelFiction-Ja_En-100k}{Hugging Face} & Model/dataset \\
Background erase network & \href{https://huggingface.co/PramaLLC/BEN}{Hugging Face} & Model \\
Omnivision-968M & \href{https://huggingface.co/NexaAIDev/omnivision-968M}{Hugging Face}, \href{https://nexa.ai/blogs/omni-vision}{Blogpost} & Model \\
Parler TTS v1 & \href{https://github.com/huggingface/parler-tts}{GitHub}, \href{https://huggingface.co/collections/parler-tts/parler-tts-fully-open-source-high-quality-tts-66164ad285ba03e8ffde214c}{Hugging Face} & Model \\
Click3-Android agent & \href{https://github.com/BandarLabs/clickclickclick}{GitHub} & Inference/mobile \\
PocketPal &  & Inference/mobile \\
GGUF merged & \href{https://github.com/philpax/ggml/blob/gguf-spec/docs/gguf.md}{GitHub} & Inference/extension \\
GGUF security advisory &  & Inference/extension \\
Llama 3 GGUF conversion bug &  & Inference/extension \\
GGUF development for llama.cpp & \href{https://github.com/ggerganov/llama.cpp/pull/2398#issuecomment-1682404719}{GitHub} & Inference/extension \\
Dangers of malicious GGUF files &  & Inference/extension \\
I made a little dead internet & \href{https://github.com/Sebby37/Dead-Internet?tab=readme-ov-file}{GitHub} & Inference \\
Optimize Whisper for fast inference &  & Inference \\
Success with a local voice chat & \href{https://github.com/dkjroot/iris-llm/tree/prototypes}{GitHub} & Inference \\
Copilot replacement & \href{https://github.com/ex3ndr/llama-coder}{GitHub} & Inference \\
KoboldCpp 1.67 released & \href{https://github.com/LostRuins/koboldcpp/releases/latest}{GitHub} & Inference \\
LLM scraper turns webpage into structured data & \href{https://github.com/mishushakov/llm-scraper/}{GitHub} & Inference \\
llama.cpp breaking change & \href{https://github.com/ggerganov/llama.cpp/pull/1508}{GitHub} & Inference \\
FastApply-open source Cursor & \href{https://github.com/kortix-ai/fast-apply}{GitHub}, \href{https://huggingface.co/Kortix/FastApply-7B-v1.0}{Hugging Face} & Inference \\
llama.cpp --logit-bias flag &  & Inference \\
Qwen2-Audio & \href{https://github.com/NexaAI/nexa-sdk}{GitHub}, \href{https://huggingface.co/NexaAIDev/Qwen2-Audio-7B-GGUF}{Hugging Face} & Inference \\
Vast.ai cloud inferencing guide &  & Inference \\
voicechat2 & \href{https://github.com/lhl/voicechat2}{GitHub} & Inference \\
LLM4SQL & \href{https://github.com/gd03champ/llm4sql}{GitHub} & Inference \\
NexaAI—Ollama alternative & \href{https://github.com/NexaAI/nexa-sdk}{GitHub} & Inference \\
\end{longtable}

\section{Topic Model Output}\label{app:topic}

% Auto-generated LaTeX longtable for topics.csv. Requires \usepackage{longtable,booktabs,array} in your preamble.
\begin{center}
\small
\begin{longtable}{p{0.04\textwidth} p{0.18\textwidth} p{0.12\textwidth} p{0.10\textwidth} p{0.42\textwidth}}
\caption{Topics generated by the topic model, with natural language representation, count, and post example}\\
\toprule
Topic & Topic Representation & Posts Selected & Total Post Count & Post Example \\
\midrule
\endfirsthead
\toprule
Topic & Topic Representation & Posts Selected & Total Post Count & Post Example \\
\midrule
\endhead
\midrule
\multicolumn{5}{r}{\textit{Continued on next page}}\\
\endfoot
\bottomrule
\endlastfoot
-1 & Language model comparison and development & 12 & 23123 & ``Llama 3.1 Discussion and Questions Megathread'' \\
1 & Open source AI development & 12 & 1781 & ``Why all AI should be open source and openly available'' \\
2 & Exploring use cases and learning resources for local LLMs & 15 & 1054 & ``What are people running local LLMs for?'' \\
3 & Speech and language models & 8 & 912 & ``Success with a local voice chat agent'' \\
4 & Prompt engineering & 7 & 905 & ``Get Llama 2 Prompt Format Right'' \\
6 & Multimodality & 3 & 727 & ``Ollama Alternative for Local Inference Across Text, Image, Audio, and Multimodal Models'' \\
7 & Cloud hosting and cost-effectiveness for LLMs & 10 & 674 & ``Have you truly replaced paid models with self hosted ollama or hugging face?'' \\
9 & Retrieval Augmented Generation (RAG) & 8 & 572 & ``Yet another RAG system - implementation details and lessons learned'' \\
11 & Quantization and model comparisons & 8 & 476 & ``I created a new benchmark to specifically test for reduction in quality due to quantization and fine-tuning…'' \\
12 & LLM Comparisons & 6 & 475 & ``LLM Pro/Serious Use Comparison/Test: From 7B to 70B vs. ChatGPT!'' \\
14 & Llama model issues and troubleshooting & 6 & 443 & ``GGUF is going to make llama.cpp much better and it's almost ready'' \\
17 & Local LLM translation & 7 & 367 & ``Local translator based on decent LLAMA model on personal computer and it works well!'' \\
18 & Story writing models & 1 & 363 & ``OpusV1 — Models for steerable story-writing and role-playing'' \\
20 & Fine-tuning with synthetic data & 5 & 333 & ``What's the best way to create a large synthetic dataset with a Local LLM?'' \\
22 & Code assistants & 5 & 311 & "Introducing Fast Apply - Replicate Cursor's Instant Apply model'' \\
25 & LLM UI Recommendations & 4 & 261 & ``What LLM frontend are you using?'' \\
26 & Web search and scraping & 5 & 224 & ``I massively updated my python program that allows local LLMs running via llama.cpp to look things up on the internet, it now fully web scrapes the most relevant results!'' \\
33 & Running models on mobile devices & 1 & 177 & ``Run Llama 3.2 3B on Phone - on iOS \& Android'' \\
34 & PDF data extraction and RAG improvement & 1 & 159 & "How we Chunk - turning PDF's into hierarchical structure for RAG'' \\
39 & Natural Language to SQL & 3 & 127 & ``I've built a project that integrates LLM with DBMS ( Finally ).'' \\
41 & Medical AI models & 2 & 117 & ``Democratizing Medical LLMs for 50 Languages'' \\
44 & Open model leaderboards & 2 & 104 & ``New interface for the Open LLM Leaderboard! Should be way more usable :)'' \\
50 & ChatGPT alternatives & 1 & 89 & ``cancelled my gpt-4 acc today, looking for a replacement.'' \\
\end{longtable}
\end{center}

\clearpage
\section{Data Flowchart}\label{app:flow}

\begin{figure}[h!]
    \centering
    \includegraphics[width=1\linewidth]{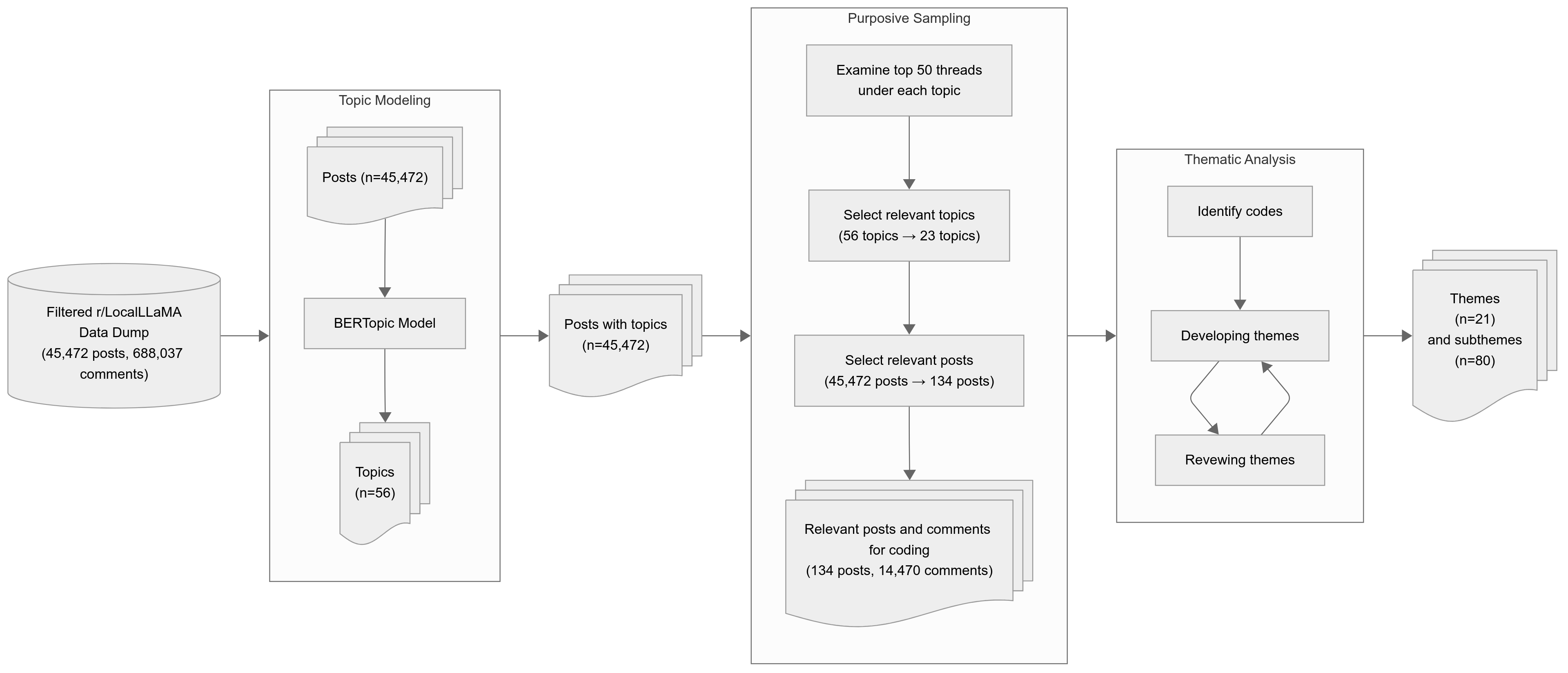}
    \caption{Flowchart of data processing, training, sampling, and thematic analysis}
    \label{fig:flowchart}
\end{figure}
\clearpage

\section{Table of Themes and Subthemes for RQ1 and RQ2}\label{app:rq12}
% Auto-generated LaTeX table. Requires \usepackage{longtable,booktabs,array} in your preamble.
\begin{center}
\small
\begin{longtable}{p{0.1\textwidth} p{0.05\textwidth} p{0.15\textwidth} p{0.07\textwidth} p{0.47\textwidth}}
\caption{Themes and subthemes that emerged from thematic coding of RQ1 data}\\
\toprule
Theme & Theme count & Subtheme & Subtheme count & Subtheme Example \\
\midrule
\endfirsthead
\toprule
Theme & Theme count & Subtheme & Subtheme count & Subtheme Example \\
\midrule
\endhead
\midrule
\multicolumn{5}{r}{\textit{Continued on next page}}\\
\endfoot
\bottomrule
\endlastfoot
Open Source & 314 & Regulation & 85 & "They can't stop open source altogether, but they can heavily stifle it by passing "AI safety" regulations...so that only big data corporations have access to quality training data...what they are really doing is pushing for expansive new censorship and surveillance regulations that are going to make it much more difficult to build and distribute open AI models." \\
 &  & Convenience vs. Openness & 61 & ``I have a copy of firefox running on my private hardware. Why is it any harder to do it with LLMs?...I go to Ollama. Can't download it for Windows. How do I start it? Go to the terminal...Honestly, I dont even need to go any further than this to know its a usability nightmare.'' \\
 &  & Terminological Rigor & 58 & ``As a researcher I am sick of the use [of] the saying open source. You are not OS unless you are completely replicable. Not a single paper since transformers has been replicable." \\
 &  & Community & 36 & ``All of this being said, willing to contribute my GPUs (2xA6000, 2x 4090) to any serious proposal to train an open source GPT-3.5 equivalent model. Curious how much interest there is.'' \\
 &  & Innovation &  & ``The RL part [of the Deepseek R1 model] has been reproduced [in a fully open source manner] already'' \\
 &  & Openness Pragmatism & 32 & "Even if Meta release the whole dataset + code, its not like everyone in their bedroom can suddenly download + modify + run it… If Meta want to call their stuff "Open Source" I don't really care, they are certainly currently greatly contributing to the OSS community. Releasing the full foundation model is in the spirit of "Open Source" in my personal opinion." \\
 &  & Resource Gap & 14 & "I am biased as one of the llama.cpp developers but my opinion is that there is more than enough work going towards training models and not enough effort going towards improving the surrounding software ecosystem. In llama.cpp/GGML for example I feel like we're chronically understaffed." \\
 &  & Democratization & 9 & ``I agree that OpenAI is trying to monopolize LLM tech, and that open weight models are the competitors' response to make it difficult to monopolize LLM tech. Imagine if there was no llama, no qwen, and so on and we had no choice but to use "OpenAI", what a nightmare.'' \\
Privacy & 129 & Sensitive Data & 89 & ``It's easier than getting a BAA for HIPAA compliance. Also much much faster for automated integration testing.'' \\
 &  & Convenience vs. Privacy & 22 & ``Local llm's won't get popular as casual users don't care about privacy but care about convenience. Companies training LLM models are losing money and won't be able to flip into profitability due to competition - most of the companies you see posting on Huggingface will be defunct in a few years.'' \\
 &  & Private nature of conversational data & 18 & "The particularly area that I am involved in uses LLMs as a way of trying to help a patient talk their way through a solution. This isn't a blind process, but a tool that a therapist or clinical technician can use to help a patient better. Sometimes patients will want to talk but the timing will not be opportunistic so being able to talk to this LLM mono to express their feelings can often help in the healing process and move them forward in their lives.'' \\
Tinkerability & 284 & Open tool building & 219 & "I have a program, where the local LLM reads my RSS feed for me and then re-orders it based on my interests before I open my RSS reader in the morning'' \\
 &  & Experimentation and learning & 46 & ``My end goals are basically [to] tinker with these things while they're still new so that I can know how it works under the hood, so that when AI becomes more mainstream I'll have a leg up, since my field (development) feels like it's right there with artists on the chopping block when AI gets better.'' \\
 &  & Affordances of openness & 19 & ``With kind of 'breakthroughs' nearly daily released or announced (e.g. see today's 'SqueezeLLM') I can hardly feel anything remotely like 'novelty wearing off'" \\
Cost & 144 & Economies of scale & 54 & "If you ran that laptop absolutely non-stop since the day it came out...[i]t would take 6 years of running your M4 [Apple's consumer hardware] at its absolute limit on a small (13B) LLM to cover the cost of the M4 when comparing it to using 4o [OpenAI's model]." \\
 &  & Small models good enough & 26 & ``Small models are sufficient for the "regex on crack" use case, which IMHO is one of the most compelling uses for LLMs, something I've wanted for decades. Also good for classification and formatting. The grammar and context awareness was pure sci-fi just 10 years ago.'' \\
 &  & Mixed use & 21 & ``I use local models for things that I feel are sensitive and I do not want to be uploaded to the cloud and I use chatgbt and others when I just want quick answers for not important things.'' \\
 &  & Local is free & 27 & "I already have the hardware, so it's cheaper to play/research with the smaller models that fit my hardware than rent computing power from a corporation." \\
 &  & Model optimization strategies & 20 & ``Rag help with that a lot ("that" refers to the small context window problem in original comment)'' \\
Data & 66 & Data licensing & 27 & ``It absolutely does not automatically warant everything to be open source, unless there is specific content included to enforce this. The GNU/GPL license is such an example. However, just learning from reading GNU/GLP licensed code, does not mean you have to publish your future code for free.'' \\
 &  & Value of data labor & 10 & ``Had they only scraped data that wes annotated to be accessible for training, public domain and explicitly permitted data I'd buy your argument but it's essentially looting the commons with the current state of things.'' \\
 &  & Not worth commoditization & 11 & ``To me the question is more why we agreed to waive our personal data to all these companies feeding closed source AI in the first place and whether we did it / keep doing it for an appropriate consideration.'' \\
 &  & Multilinguality & 7 & ``Multilingual stuff would be great because there are currently like one open weight model (which is like over 300B params..) that is good at my language (Finnish). All the other open models, Gemma, Llama, Qwen, Mistral and whatever mainly just support English or Chinese.'' \\
 &  & Right to keep data open & 6 & ``I think if they don't provide data to open datasets they're a bunch of shitbags. I'm largely abandoning reddit - I hope you guys jump to a new community.'' \\
 &  & Commoditize in exchange for platform & 6 & ``reddit doesn't cost me anything. Thus, I am the product. I know this. I accept this.'' \\
Prof & 114 & Openness as a business strategy & 94 & ``...llama.cpp, Vicuna, WizardLM, Unsloth, ExLlama, and a myriad of other projects now exist which Meta didn't have to create themselves. It's not (just) out of the goodness of their hearts, they are getting a financial benefit and are actively saving money by letting their competitors use their work.'' \\
 &  & Open models for business & 18 & ``Ye so Pro makes training even faster from 5X to 28X ish faster, supports multi GPU training. Max further speeds it up to 31x, but the difference is Max makes it possible to work on Intel, AMD GPUs, and supports full finetuning and training.'' \\
 &  & Economics of open model development & 2 & ``If we mandated that all AI should be open source, there would be less of it, because now that investment cash goes elsewhere and selfish persons pursue other careers…'' \\
Roleplay & 42 & Roleplay good & 24 & ``as woman I prefer erp then porn, porn is too male gazey, not enough emotion and not enough entertainment, plus some of the positions they put the women in look like they hurt…'' \\
 &  & Sexual desires drives progress & 9 & ``LLM ERP weirds me out… But, in all seriousness, porn is always an early adopter for new tech and a huge driver of innovation. If it pushes the technology forward, I ain't gonna try to stop `em.'' \\
 &  & Roleplay bad due to quality & 5 & ``Long-time ERPer here. I never played with an AI because part of the point why I ERP instead of writing erotica is that I'm teaming up with another human. The fact I know it's an LLM and I know what it's doing, how it never understands or cares for anything breaks the magic for me.'' \\
 &  & Roleplay bad due to risk & 2 & ``At the same time, RP is NOT a good substitute to actual socialization and relying on it could potentially warp your view of people (\**cough\** assistant bias \**cough\**).'' \\
 &  & Roleplay neutral & 2 & ``I don't see it as much different from playing a video game (or most other forms of entertainment for that matter), when used responsibly. A relative term I know. But hey 20 years ago a ton of adults thought other adults playing video games was cringy and unhealthy... as they sit on the couch to watch stuff they don't even care about on TV for hours.'' \\
Censorship & 34 & Models shouldn't be censored & 26 & ``Uncensored allows for better reasoning. Trying to block the model from producing nsfw outputs makes the model worse because humans naturally produce nsfw outputs under the right circumstance (think flirting, dating, romance)…'' \\
 &  & Uncensored models depict more accurate representations of humans & 8 & "Uncensored LLMs are a reflection of the human experience, in both good and bad. You could use uncensored LLMs to gain insight into how people think — not as individuals, but as a statistical representation. And the results are pretty dark." \\
Independence & 28 & Avoiding vendor lock-in & 12 & "I certainly don't want to be sending all my codes and documents off to some American company that's getting rich off my data." \\
 &  & Permanent ownership & 9 & ``You can do much more with a couple of 3090 than just llm, you open a rabbit hole into machine learning. It's a lot of learning but I find it worth it. Openai subscription just gives you temporary access to a model you don't know how and why it's working neither which bias and limitations it has.'' \\
 &  & Anti profit-first & 7 & "I can't trust profit oriented companies not enough about AI in general. They do all for making money and that didn't match for making good AI for users. "good" is a term that companies always bend for their profit." \\
Robustness & 28 & Reliability & 24 & "I used to be an early ChatGPT user since their public beta, but as the time went by, they broke my workflows countless times, doing updates without my consent - some prompts that used to give mostly reliable result started to give either wrong results or no results at all." \\
 &  & Transparency & 4 & "I want to know that the llm isn't augmented behind the API. That way you can attribute every performance increase to either a better architecture or training regime or better data." \\
Performance & 23 & Open AI is better than proprietary AI & 12 & ``I've had multiple instances now where Sky-T1 has given me code that works *almost* perfectly on the first run, and definitely gets it after explaining what's wrong with it. Whereas O1 gives me buggy code I have to give it the errors or explain the problems 2-3+ times. Anyone else had this experience?'' \\
 &  & Proprietary AI is better than open AI & 11 & ``No way. I am still waiting for the day local models will be able to compete with state of the art like the 200\$ OpenAI plan. As of now, no model can. No, not even DeepSeek.'' \\
Risk & 20 & Perceived risk high & 14 & ``The "long-term relationships" these companies sell seem especially isolating. Not to mention the dystopian data harvesting likely happening. Hard not to judge the people getting rich off this.'' \\
 &  & Perceived risk low & 14 & ``Does the absence of an LLM prevent people from scamming others? Do LLM's promote scamming others?...Just because one AI can be used to scam people, doesn't necessarily mean all AI technology should be regulated the same way.'' \\
Localness & 18 & Offline use & 18 & "It's somehow like having an internet connection even offline to find answers to smaller/medium question (mostly programming)." \\
Politics & 13 & Role of democracy in AI development & 13 & ``That fact he got voted in with an election makes this all kind of dumb. Please let me know when Xi s next election is? Not having to be politically accountable is a lot different than saying a lot of dumb stuff on truth social'' \\
\end{longtable}
\end{center}
\clearpage
\section{Table of Themes and Subthemes for RQ3}\label{app:rq3}

\begin{longtable}{p{0.1\textwidth} p{0.05\textwidth} p{0.2\textwidth} p{0.07\textwidth} p{0.4\textwidth}}
\caption{Themes and subthemes that emerged from thematic coding of RQ3 data} \\
\toprule
Theme & Theme count & Subtheme & Subtheme count & Subtheme Example \\
\midrule
\endfirsthead
\toprule
Theme & Theme count & Subtheme & Subtheme count & Subtheme Example \\
\midrule
\endhead
\midrule
\multicolumn{5}{r}{Continued on next page} \\
\midrule
\endfoot
\bottomrule
\endlastfoot
Conditions &  & Addressing a shared and in-demand problem of the community & 28 & “Hey, like many of you folks, I also couldn't wait to try llama 3.2 on my phone. So added Llama 3.2 3B (Q4\textbackslash \_K\textbackslash \_M GGUF) to PocketPal's list of default models, as soon as I saw this [post] that GGUFs are available!” \\
 &  & Peer education & 14 & “We've spent a lot of time building new techniques for parsing and searching PDFs. They've lead to a significant improvement in our RAG search and I wanted to share what we've learned.” \\
 &  & Asking the community to share their projects to be inspired & 5 & “Which RAG projects have you developed that you're most proud of? I've recently started making RAG apps using ollama and python, while they work I wouldn't call them perfect. So I'd love to see what a good RAG application has under the hood…” \\
 &  & Desiring post to inspire others & 3 & “So, I don't really have much to say except sharing my progress and thanks, but I hope it might inspire some people to share what they've achieved…” \\
 &  & Giving back to the community & 3 & "I've been part of this community for a while and have gained a lot from your insights and discussions. Today, I'm excited to share a project I've been working on called AgentSearch…” \\
 &  & Providing entertainment to the community through one’s project & 1 & “Ever wanted to surf the internet, but nothing is made by people and it's kinda janky? No? Too bad I made it anyways! You can find it [here on my Github](https://github.com/Sebby37/Dead-Internet), instructions in README. Every page is LLM-generated, even the search results page! Have fun surfing the “net”!" \\
 &  & Desiring feedback from the community & 19 & “As always, your feedback is super valuable! Feel free to share your thoughts or report any bugs/issues via GitHub” \\
Action: Problem articulation &   & Help on whether model fits my computer and/or how fast it will run & 19 & What I am wondering is if you quantize a 32B distilled model to 1.58 bits in this same method, will it perform equally well, better or worse and faster or slower than a 14B distilled 4bit AWQ? \\
 &  & Asking for tutorials on how to use stack/model & 23 & Can someone point me in the direction of a step by step install guide for the 7b uncensored? \\
 &  & Asking for additional details on what the poster provided & 100 & Any chance you can share your training code? I want to fine-tune it using PEFT, but new to training LLMs. \\
 &  & Asking for plans in the future for given project & 9 & Hey, Wolfram. Are you planning to use new test on new models(llama 3.1, two new mistral models, and many more)? \\
 &  & Asking for others' experience with the artifact & 27 & i will really like to see some solid examples demoing this if someone can provide. \\
 &  & Asking for the source code/model if it does not exist & 10 & Can you please release code needed to perform this manually for models where you didn't upload the quants? I'm planning to finetune Qwen2 VL 72B with QLoRA and I would also like to see how this affects text only llm's I've been using qlora on. \\
 &  & Asking whether a benchmark for the project exists & 7 & What engine are you using in the backend to run it? Or how did you run the benchmarks? I would like to try to reproduce the results and compare it to other quants. \\
Action: Feedback solicitation & 51 & Asking for follow-ups & 15 & Hi! If it is convenient for you, could you please provide the example? This would help us improve the model! \\
 &  & Asking for the OP's opinions on something & 3 & Out of curiosity since both models have been out for a while, what is your impression of Mistral 7B OpenOrca compared to OpenHermes? \\
 &  & Asking for how the project compares to something else & 14 & How does it compare to mistral-large EXL2 quants? \\
Action: Solution circulation & 587 & Suggestion for future release & 87 & I think the biggest help right now would be BitNet models. \textasciitilde 8x model size reduction as well as removing matrix multiplication opens a whole new area for optimization. What's available right now seems promising, but the big question is how well they scale (past 3B parameters and 300B tokens). \\
 &  & Bug report/New GitHub Issue & 24 & There's a bug somewhere for sure. Either in your benchmark, or in the loader, or in the quantization code. \\
 &  & Offering help & 9 & I am fine-tuning yi-34b on 24gb 3090 ti with ctx size 1200 using axolotl. If you want some tips and tricks with it I can help you to get up to what I am getting. \\
 &  & Sharing papers & 13 & Daniel you might want to look at this paper, section 4.3. Compression Scheme \& Kernel Co-design: Which allows a further drastic memory reduction by compressing the quantized weights. \\
 &  & Sharing repos/other projects & 63 & Their github explains what their motivation / manipulations are here: [link]. So, looks like they tweaked the WizardLM conversations to make them more conversational in nature rather than instructional, then mixed in the Vicuna bits. \\
 &  & Sharing blogs/notebooks/documentation & 22 & https://unsloth.ai/introducing has more deets on the manual autograd methods and Triton kernels. + other coding tricks like inplace operations, reduced memory movements, etc. \\
 &  & Sharing technical knowledge & 44 & I found CUDA kernels for non jitted code to be faster - ie if you run CUDA kernels only once or twice since there's a JIT compiling cost via Triton. In general, CUDA and Triton are equal in terms of speed - Triton more so since you can try out more hypotheses. \\
 &  & Sharing personal experience & 101 & u/YearZero this was the best 7b model I've found in my personal testing, you should see how this stacks up against other 13b models! \\
 &  & Sharing requested details on the project & 73 & Yes its a model format by itself. Unlike koboldcpp/llama.cpp, exllama is for gpu only, no cpu offloading pretty much. It's pretty much the fastest model loader you can get. \\
 &  & Redirecting questions to other members of the community & 9 & LoneStriker takes quant requests! Try creating a new discussion in LZLV's Hugging Face page requesting for new quants and @ them in your post. \\
 &  & Accountability & 26 & It's best practice to require opt-in to analytics when user data is concerned. This variable should default to False. \\
Action: Social reinforcement &  &  &  & This is probably one of the most significant pushes to the open source AI community. Thanks, guys. Cheers \\
Consequence: Iterative and interdependent development &  & Adding additional datapoints to OP's post & 4 & Benched qwen2.5-14b-instruct-q4\_k\_m on computer science using the same method as OP, but with LM Studio instead of Ollama. Result using the default config: Total, 268/410, 65.37\%, Random Guess Attempts, 0/410, 0.00\%, Correct Random Guesses, division by zero error, Adjusted Score Without Random Guesses, 268/410, 65.37\% \\
 &  & Making hypotheses based on the OP's empirical evidence & 1 & Curious if by training it in Dutch it would become a better translation agent between that and English, and if so how to finetune for other languages. \\
 &  & OP extending their work based on feedback & 9 & Oh wow, thanks for the heads-up. I looked at that literally just a few days ago and it wasn't there yet, just updated right now, and there it is. I foresee some new comparisons coming up... ;) \\
 &  & OP asking for a specific commenter's contribution & 2 & Of course, feel free to contribute on git and/or produce datasets or models using our code. The code is pretty young so we can use some help maturing the scripts. \\
 &  & Another project being mentioned as having relevance and synergy & 1 & You can push it to under \$0.50 with my OSS package Unsloth if you're finetuning more models! It makes finetuning via QLoRA **2.2x faster** and use 62\% less memory, so you can wait less, pay less and increase the batch size! If you want to collab on finetuning more on other languages, more than happy to help! \\
 &  & Compute donation/funding to sustain an open source project & 5 & I should be able to provide access to a few instances each with 8x RTX 3090. Please reach out via DM to me should this be of interest :) \\
 &  & GitHub/other updates to the original thread & 53 & It's in the "experimental" branch, but there's a bug in the most recent commit, so you should hold off at least an hour or so. \\
 &  & Collaborative ideation & 58 & One possibility, is to organize experts to 64 groups (4 experts in each group) and collapse each group to a single experts, getting 64 experts. This adds quite a lot of complexity though, and also there is a question on what criteria experts should be put in a single group (I guess could be done randomly as the most simple approach). \\
\end{longtable}

\clearpage
\section{Qualitative Coding Tool}\label{app:tool}

\begin{figure}[h]
  \centering
  \includegraphics[width=0.75\linewidth]{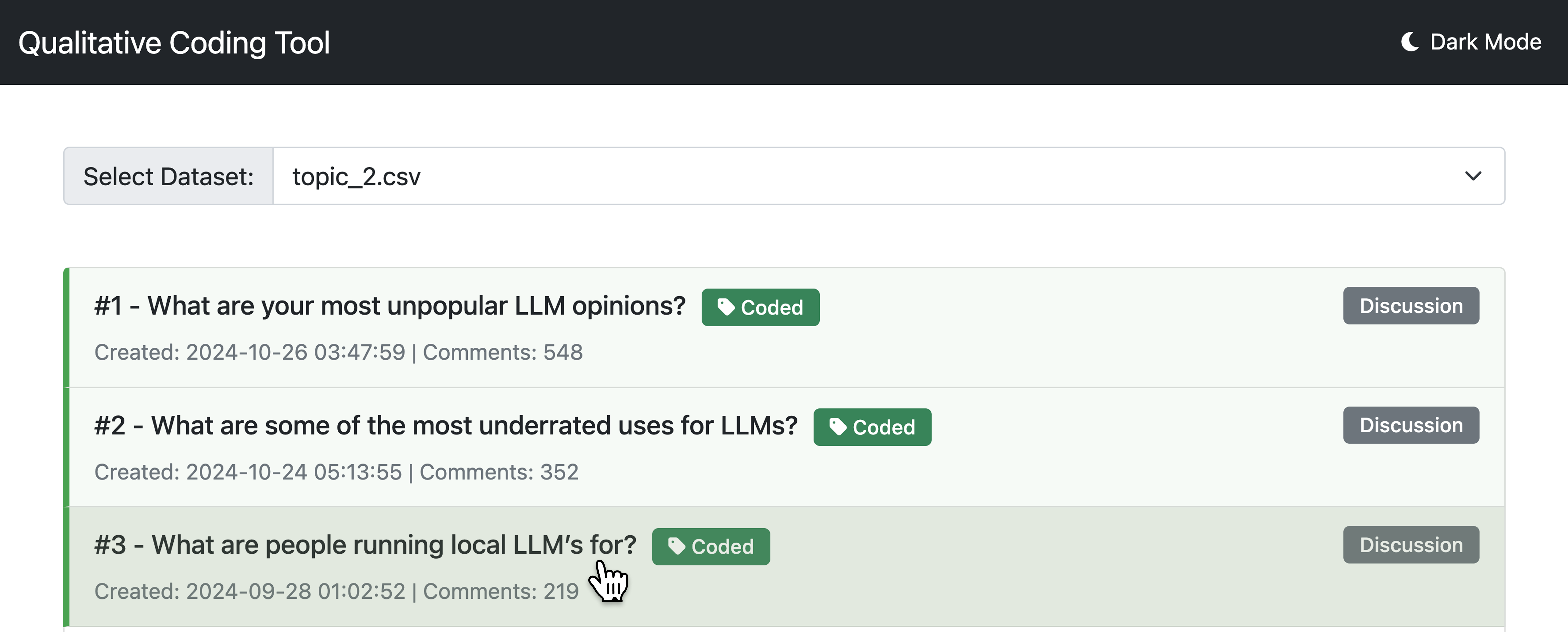}
  \caption{Index page of the Qualitative Coding Tool}
  \Description{The index page of the Qualitative Coding Tool, including a data selection drop-down, and a list of posts under the selected dataset.}
  \label{fig:topic}
\end{figure}

\begin{figure}[h]
  \centering
  \includegraphics[width=0.9\linewidth]{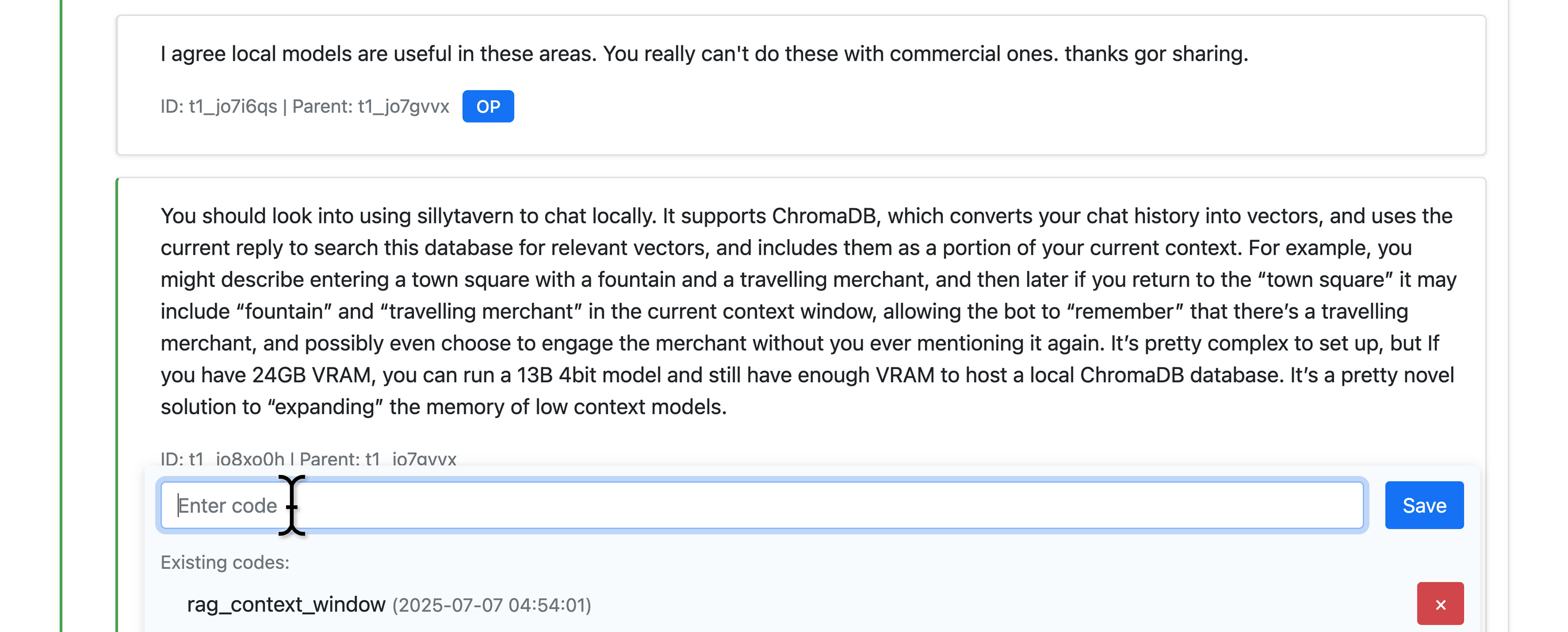}
  \caption{Qualitative coding interface for threads of comments.}
  \Description{The interface for qualitative coding a thread of comments, including the comment itself and a hover box below it to enter codes.}
  \label{fig:coding}
\end{figure}

For thematic analysis, we wanted to code comments under each post in a way that preserves the thread-format that is typical of a discussion forum's user interface. Since we did not find a tool suitable for our needs, we developed a web qualitative coding tool that converts a CSV file of discussion forum data to threads with a text box under each comment for storing codes (See Fig. \ref{fig:topic} and Fig. \ref{fig:coding}). These codes are stored on a database, and can be exported when the researcher has completed the coding process. The code\footnote[1]{\url{https://github.com/kweel/qual-coder-web}} for this tool is provided for researchers to use and improve as needed.

\clearpage
\section{Endmatter}

\subsection{Generative AI Usage Statement}

ChatGPT---the most up-to-date web version from chatgpt.com---and Gemini---the most up-to-date web version from gemini.google.com---were used for the following tasks:

\begin{itemize}
    \item Generating LaTeX tables from CSV files
    \item Pointing out sections that could be shortened in length to meet the page limit
    \item Suggesting ways to rewrite sections for better readability, like condensing or dividing long sentences.
\end{itemize}

All generations were thoroughly checked by the authors before their adaptations into the manuscript.

\subsection{Limitations and Ethical Considerations}

To mitigate some of the known issues and factors around using Reddit as a data source \cite{PROFERES_2021}, we have considered the following:

\begin{itemize}
    \item  We recognize that public data does not mean ethical clearance, and that users may perceive their posts as semi-private within online communities \cite{beninger2017social}. Therefore, we refrained from including personally identifiable or sensitive content and ensure all quoted material is anonymized to prevent direct re-identification, while striving to maintain the semantic integrity of the user's lived experience.
    \item  We recognize that archival snapshots do not perfectly mirror its up-to-date data \cite{gaffney_2018}. We made the best effort to cross-check content using different sources, get full coverage of the data, and remove content deleted by the user in any one of the cross-checked datasets.
    \item This study is limited by the demographic biases of Reddit, which tends to overrepresent younger and male users. Furthermore, it is highly difficult to get comprehensive data on the demographics of Reddit and its subreddits \cite{reddit_dem}. We also note that user flairs were not used consistently, with 99\% of unique users who have either posted or commented on r/LocalLLaMA having none, and thus the lack of systematic demographic data as a study limitation. This limitation means that our findings may not reflect the full picture of perspectives and practices with open models in the wild. Future work may explore triangulating different data sources to extend our work.
    \item  We hope that the findings of this paper will be informative for the community, so we plan to post it to r/LocalLLaMA pending publication, with a specialized abstract detailing some of our findings as well an acknowledgment of the value of their data to our research.
\end{itemize}

\end{document}